\documentclass[aps,prb,twocolumn,showpacs]{revtex4}
\usepackage{graphicx,epsf}
\usepackage{amsmath}

\newcommand{\up}{\uparrow}
\newcommand{\dn}{\downarrow}

\newcommand{\afs}{AF$_{\rm S}$}
\newcommand{\afl}{AF$_{\rm L}$}

\newcommand{\CeNi}{CeNi$_2$Ge$_2$}
\newcommand{\CeAu}{CeCu$_{6-x}$Au$_x$}

\newcommand{\YbRhSi}{YbRh$_2$Si$_2$}

\begin{document}

\title{
From itinerant to local-moment antiferromagnetism in Kondo lattices:\\
Adiabatic continuity vs. quantum phase transitions
}

\author{Matthias Vojta}
\affiliation{Institut f\"ur Theoretische Physik, Universit\"at zu K\"oln,
Z\"ulpicher Str. 77, 50937 K\"oln, Germany}
\date{Sep 1, 2008}

\begin{abstract}
Motivated by both experimental and theoretical activities,
we discuss the fate of Kondo screening and possible quantum phase transitions
in antiferromagnetically ordered phases of Kondo lattices.
While transitions with topological changes of the Fermi surface may occur,
we demonstrate that an entirely continuous evolution
from itinerant to local-moment antiferromagnetism
(i.e. from strong to negligible Kondo screening)
is possible as well.
This situation is in contrast to that in a non-symmetry-broken situation
where a quantum phase transition towards an exotic
metallic spin-liquid state necessarily accompanies the disappearance
of Kondo screening.
We discuss criteria for the existence of topological transitions in the
antiferromagnetic phase, as well as implications for theoretical scenarios and for
current experiments.
\end{abstract}

\maketitle


\section{Introduction}

Quantum criticality in heavy-fermion metals is an active topic
in current condensed-matter research.
Much work has focused on the nature of the quantum phase transition (QPT)
between a paramagnetic heavy Fermi liquid (FL) and an
antiferromagnetic (AF) metal.\cite{doniach,hvl}
Various experimental data appear to be inconsistent with
the theoretical predictions for a spin-density wave (SDW)
transition in a metal.\cite{hertz,millis,moriya}
This has prompted proposals about a different transition
scenario, where the Kondo effect breaks down at the
antiferromagnetic quantum critical point (QCP).
Then the heavy quasiparticles of FL,
formed from conduction ($c$) and local-moment ($f$)
electrons, disintegrate at the QCP.

Such a Kondo-breakdown transition involves degrees of
freedom other than the antiferromagnetic fluctuations
at the ordering wavevector $\vec Q$, and
different theoretical descriptions have been
proposed.\cite{coleman,si,senthil}
Si {\em et al.}\cite{si} have employed an extension of
dynamical mean-field theory (DMFT) to argue that magnetic fluctuations
in two space dimensions render the Kondo effect critical {\em at}
the AF QCP.
In contrast, the scenario of Senthil {\em et al.}\cite{senthil,senthil2,senthil3,senthil4}
is centered around a Kondo-breakdown transition between two {\em paramagnetic} states:
a heavy Fermi liquid with Kondo screening and
a so-called fractionalized Fermi liquid (FL$^\ast$) without Kondo screening.
In FL$^\ast$, frustration and/or strong quantum fluctuations preclude
magnetic long-range order of the $f$ moments,
which instead form an exotic spin liquid,\cite{frust_sl}
decoupled from the conduction electrons.
A sharp distinction between FL and FL$^\ast$ is in the
Fermi volume (assuming one $f$ electron per unit cell in the paramagnet):
FL has a ``large'' Fermi volume including the $f$ electrons,
whereas the Fermi volume of FL$^\ast$ is ``small'', i.e.,
only determined by the $c$ electrons.
Loosely speaking, the $f$ electrons may be called ``itinerant'' in FL
and ``localized'' in FL$^\ast$.
In this scenario, antiferromagnetism may occur as a secondary instability of FL$^\ast$,
such that a conventional AF phase is reached via the Kondo-breakdown
transition.\cite{senthil3,senthil4}

The possibility of having distinct quantum critical points between FL and AF
raises the question whether distinct AF {\em phases}
(i.e. with ``itinerant'' or ``localized'' $f$ electrons)
may be discriminated.
Consider commensurate order with an even number $N$ of sites in the
unit cell (other cases will be discussed briefly towards the end of the paper).
Then, the size of the Brillouin zone in the AF phase is reduced by a factor $N$
compared to the paramagnetic phase, and as a result
``large'' and ``small'' Fermi volume are no longer distinct.
However, it has been proposed that the situations differ
w.r.t. their Fermi-surface topology.\cite{senthil2,si_global,yama,ogata}
Two phases, \afl\ and \afs, corresponding to itinerant and local-moment
antiferromagnetism, respectively, have been introduced,\cite{si_global,yama}
with the notion that they are separated by one or more
quantum phase transitions.
In this context, it has been suggested that in \afs\ Kondo screening is absent.

In this paper, we argue that such a sharp distinction between \afl\ and \afs\ does not
exist.
To this end, we invoke continuity arguments between Kondo and weakly interacting
electron models, and demonstrate the possibility of a continuous Fermi surface (FS) evolution
between the situations of itinerant and local-moment antiferromagnetism.
Kondo screening hence disappears smoothly in the AF phase:
Technically, a line of renormalization-group fixed points emerges,
describing AF polarized Fermi liquids.
This finding does not contradict microscopic calculations\cite{ogata,assaad}
which find a topological Lifshitz transition inside the antiferromagnetic phase,
but shows that such a transition is not connected to
a breakdown of the Kondo effect.
We also discuss conditions for the occurrence of FS-topology-changing
transitions.

We note that many experimental criteria for Kondo screening,
e.g. the existence of a maximum in the
resistivity $\rho(T)$ at the coherence temperature,
do not provide a sharp distinction between itinerant and local-moment AF.
This means e.g. that the maximum in $\rho(T)$ will be gradually washed out and
disappear when tuning from itinerant to local-moment AF.

The remainder of this paper is organized as follows:
In Sec.~\ref{sec:gen} we start by characterizing
ground states of Kondo lattices,
and we discuss the adiabatic continuity to phases of weakly
interacting electrons.
In Sec.~\ref{sec:mf} we give an explicit example for a continuous
Fermi-surface evolution in a mean-field Kondo lattice model,
smoothly connecting the situations of ``itinerant'' and ``localized''
$f$ electrons in the presence of commensurate antiferromagnetism.
We then re-formulate our findings in the languages of
renormalization group (Sec.~\ref{sec:rg}) and
slave-particle theory coupled to gauge fields (Sec.~\ref{sec:gauge}),
establishing connections to earlier work.
We close with remarks on recent theoretical and experimental results.
The detailed discussion of band structures, of criteria for topological transitions
and of the interesting case of incommensurate magnetic order
are relegated to the appendices.


\section{General considerations}
\label{sec:gen}

Consider a Kondo lattice model in $d$ spatial dimensions with a unit cell
containing one $c$ and $f$ orbital each,
\begin{equation}
\label{eq:KLM}
\mathcal{H}_{\text{KLM}}=\sum_{{\vec k}\sigma} (\epsilon_{\vec k}-\mu) c_{{\vec k}
\sigma}^{\dagger}c_{{\vec k} \sigma} +
J \sum_{i} {\vec S}_i \cdot {\vec s}_i,
\end{equation}
where the chemical potential $\mu$ controls the filling $n_c$ of the
conduction ($c$) band with dispersion $\epsilon_{\vec k}$,
and ${\vec s}_{i}= \sum_{\sigma \sigma'}
c_{i\sigma}^{\dagger} {\vec \tau}_{\sigma \sigma'} c_{i\sigma'} / 2$
is the conduction electron spin density on site $i$.
Sometimes it is useful to explicitly include a Heisenberg-type exchange
interaction between the $f$ electron local moments ${\vec S}_i$,
${\cal H}_{\rm I} = \sum_{ij} I_{ij} {\vec S}_i \cdot {\vec S}_j$,
which may originate from superexchange (or RKKY) interactions.\cite{rkkyfoot}

\subsection{Phases}

If Kondo screening of the local moments ${\vec S}_i$ dominates over inter-moment interactions,
then a heavy Fermi liquid (FL) results.
The Fermi volume is ``large'', i.e., includes the local-moment electrons,
${\cal V}_{\rm FL} = K_d (n_{\rm tot}\,{\rm mod}\,2)$ with $n_{\rm tot}=n_c+n_f = n_c+1$,
in agreement with Luttinger's theorem.\cite{oshi2}
Here, the factor of two accounts for the spin degeneracy of the bands,
and $K_d = (2\pi)^d/(2 v_0)$ is a phase space factor,
with $v_0$ the unit cell volume.

Kondo screening may break down due to competing exchange interactions
among the local $f$ moments.
If the local-moment magnetism is dominated by geometric frustration
or strong quantum fluctuations,\cite{frust_sl} the $f$ moment subsystem may form a
paramagnetic spin liquid without broken symmetries, only weakly interacting
with the $c$ electrons.\cite{senthil}
The resulting FL$^\ast$ phase is necessarily exotic,
as it features a ``small'' Fermi volume
${\cal V}_{\rm FL^\ast} = K_d (n_c\,{\rm mod}\,2)$
violating Luttinger's theorem.
As discussed in Ref.~\onlinecite{senthil}, the low-energy excitations of the
fractionalized spin liquid account for the Luttinger violation.
Thus, the Fermi volume provides a sharp distinction between FL and FL$^\ast$.

If the heavy FL phase undergoes a standard SDW transition,
we obtain a conventional metallic AF phase -- often denoted as ``itinerant''
antiferromagnet -- with Fermi-liquid properties.
For the simplest case of collinear commensurate antiferromagnetism
with an even number $N$ of sites in the AF unit cell,
the spin degeneracy of the bands is preserved (see Appendix \ref{app:band}),
and the onset of AF order simply implies
a ``backfolding'' of the bands into the AF Brillouin zone.
This results in a Fermi volume
${\cal V}_{\rm AF} = K'_d (N n_{\rm tot}\,{\rm mod}\,2)$ with $K'_d = K_d/N$.
This value of ${\cal V}_{\rm AF}$ equals $K'_d (N n_c\,{\rm mod}\,2)$,
i.e., the distinction between ``large'' and ``small'' Fermi volume is lost.

On the other hand, one may consider a ``local-moment'' AF phase.
In the language of the Kondo model, this is obtained by forming
a local-moment antiferromagnet from the $f$ spins,
and then switching on a weak Kondo coupling to the $c$ electrons.
Importantly, this phase has Fermi-liquid properties as well.

Paranthetically, we note that a distinct magnetic phase is obtained
by the onset of magnetic order in the fractionalized FL$^\ast$ phase.\cite{senthil2}
This exotic AF$^\ast$ phase, characterized by topological order,
is not of interest for the body of the paper,
but will be briefly discussed in Sec.~\ref{sec:gauge}.

\subsection{Adiabatic continuity and weakly-interacting electron states}
\label{sec:adiabatic}

The FL phase of a Kondo lattice is adiabatically connected to the
non-interacting limit of the corresponding Anderson lattice model,
\begin{eqnarray}
\label{eq:ALM}
\mathcal{H}_{\text{ALM}}&=&\sum_{{\vec k}\sigma}  (\epsilon_{\vec k}-\mu) c_{{\vec k}
\sigma}^{\dagger}c_{{\vec k} \sigma} +\sum_{{\vec k} \sigma}
(\epsilon_f-\mu) f_{{\vec k} \sigma}^{\dagger} f_{{\vec k} \sigma} \\
&+&V \sum_{{\vec k}\sigma}
(f_{{\vec k} \sigma}^{\dagger} c_{{\vec k} \sigma} + c_{{\vec k} \sigma}^{\dagger} f_{{\vec k}
\sigma}) + U \sum_i n_{f,i\uparrow} n_{f,i\downarrow}
\nonumber
\end{eqnarray}
in standard notation.
While this Anderson model can be mapped to the Kondo model in the
Kondo limit,
$V\to\infty$, $U\to\infty$, $\epsilon_f\to-\infty$ with $V^2/\epsilon_f$ finite,
the properties of the heavy FL phase of both models can
in principle be obtained perturbatively in $U$ starting from two hybridized
non-interacting bands.\cite{hewson}
In other words, no phase transition occurs between the free-fermion situation $U=0$
and the large-$U$ Fermi liquid phase in the Anderson model.\cite{scfoot}

As the itinerant AF phase is obtained from FL by the onset of SDW order,
it is adiabatically connected to a non-interacting electron system of
two hybridized bands in the presence of a staggered magnetic
mean field.\cite{affoot,pivo}

Now we turn to the local-moment AF phase.
Consider first the system of $f$ moments alone:
This is an AF Mott insulator.
In the band picture, the $f$ band is half-filled;
after backfolding this translates into integer band filling
for an even number $N$ of sites in the AF unit cell.
(The same applies to odd $N$, but here the spin degeneracy is
lifted, see Appendix~\ref{app:band}.)
As large band gaps are induced by the AF order parameter,
non-interacting $f$ electrons in the presence of an antiferromagnetic
exchange (mean) field are insulating as well,
i.e., there is no distinction here between band and Mott insulator.
This fact is well known, e.g., for a half-filled one-band Hubbard
model, where itinerant and local-moment antiferromagnet
are continuously connected upon variation of $U$.\cite{hubb_af}
Consequently, for {\em vanishing} Kondo coupling,
the local-moment AF phase of the Kondo lattice is adiabatically
connected to a non-interacting Anderson model with
vanishing hybridization and mean-field antiferromagnetism.
Further, in such a two-band model of non-interacting electrons with
$f$ band gap, small $c$--$f$ hybridization is a marginal perturbation
as it shifts the bands, but leaves the topology of the Fermi surface
unchanged.

\subsection{Evolution of phases}

Recent theory works\cite{si_global,yama,ogata}
suggested that local-moment and itinerant antiferromagnetism in
Kondo lattices are distinct phases which
are separated by (at least) one quantum phase transition.
There are, in fact, two issues here, related to
(A) possibly different Fermi-surface topologies, and
(B) a possible breakdown of Kondo screening within the AF phase,
to be discussed in turn.

(A) The local-moment AF phase (dubbed \afs) displays a Fermi
surface with a topology inherited from the bare $c$ band,
while the itinerant AF phase (dubbed \afl) was argued to display a
different FS topology.\cite{yama}
(In the simplest models, \afl\ has a hole-like FS,
whereas \afs\ has an electron-like FS.)
If this difference in topology indeed exists, it
necessitates one or more topology-changing transitions inside
the AF phase.
Such a transition can be of Lifshitz or van-Hove type,
where a local band extremum or a saddle point crosses the Fermi level,
respectively -- both cases cause (weak) thermodynamic singularities and
lead to abrupt changes in the electron orbits as measured e.g. in the
de-Haas-van-Alphen effect.
The variational Kondo-lattice-model calculations of Ref.~\onlinecite{ogata}
indeed showed two AF phases with different FS topology;
however, in a large parameter regime,
a direct first-order transition from FL to \afs\ was found.

(B) In Ref.~\onlinecite{yama}, the \afs\ phase was argued to be
stable against a small Kondo coupling. The authors concluded that
this situation is qualitatively different from \afl, and consequently
\afl\ and \afs\ have to be separated by a
quantum phase transition associated with the breakdown of
Kondo screening
(which may coincide with the transition in FS topology).

In the following, we argue that
(A') a continuous FS evolution between itinerant and local-moment AF phases
is possible (using an explicit example, Sec.~\ref{sec:mf}), and
(B') Kondo screening disappears smoothly in the AF phase
(Sec.~\ref{sec:rg}).
Consequently, a sharp distinction between two {\em phases} \afs\ and \afl\
is not meaningful.


\section{Weakly interacting electrons: Fermi-surface evolution}
\label{sec:mf}

We now give an explicit example for a continuous
evolution from itinerant to local-moment AF
in a model of effectively non-interacting electrons.
This approach can be justified by invoking
either the equivalence of the
relevant Fermi-liquid phases to phases of non-interacting electrons
(Sec.~\ref{sec:adiabatic}),
or the well-known mean-field treatments
of the Kondo or Anderson lattice models.
For instance, for the Kondo lattice the Kondo interaction
is decoupled using a slave-boson field $b$.
At the saddle-point level, one obtains a two-band model
of non-interacting electrons, with a hybridization $b$,
hence the condensation of $b$ signals Kondo screening
(see Sec.~\ref{sec:gauge} for details).
Antiferromagnetism is obtained from a mean-field decoupling\cite{senthil2,rkkyfoot}
of the inter-moment exchange interaction
$I_{ij}$, which results in a staggered mean-field for the
$f$ electrons.

We consider the following effective Hamiltonian:
\begin{eqnarray}
{\cal H}_{\rm mf} &=&
\sum_{k\sigma} \Big[
  (\epsilon_{\vec k}-\mu)        c_{{\vec k}\sigma}^\dagger c_{{\vec k}\sigma}
 +  (\epsilon_{{\vec k}f}-\lambda) f_{{\vec k}\sigma}^\dagger f_{{\vec k}\sigma} \nonumber\\
&+& V (c_{{\vec k}\sigma}^\dagger f_{{\vec k}\sigma} + h.c.) \nonumber\\
&+& m_s \sigma \,c_{{\vec k+ \vec Q}\sigma}^\dagger c_{{\vec k}\sigma}
 +  M_s \sigma \,f_{{\vec k+\vec Q}\sigma}^\dagger f_{{\vec k}\sigma}
\Big] .
\label{hmf}
\end{eqnarray}
It consists of two bands, hybridized by $V$,
with dispersions $\epsilon_{\vec k}$ and $\epsilon_{{\vec k}f}$, respectively.
Note that, within a mean-field theory for the Kondo lattice,
$\epsilon_{{\vec k}f}$ and $\lambda$ are renormalized {\em effective}
parameters of the $f$ band, the latter playing the role of a Lagrange multiplier
fixing $n_f=1$.
Collinear antiferromagnetism is accounted for via the mean fields
$m_s$ and $M_s$ (for $c$ and $f$ electrons),
encoding the magnetic order parameter with wavevector $\vec Q$
(see Appendix~\ref{app:band} for the general case).
Typically, order will primarily arise in the $f$ electron sector,
hence we expect $|M_s|>|m_s|$.
$\vec Q$ corresponds to an $N$-site unit cell in the AF ordered state,
for even $N$ the AF bands will be spin-degenerate (Appendix~\ref{app:band}).

\begin{figure}[!t]
\epsfxsize=3.2in
\epsffile{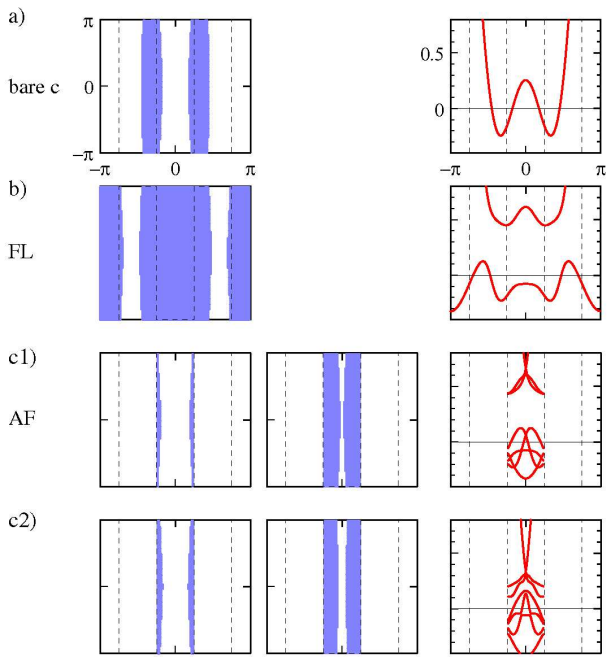}
\caption{
(Color online)
Fermi surfaces (left) and band structures (right)
for a mean-field Kondo lattice model, Eq.~\eqref{hmf},
on a 2d rectangular lattice,
with $n_c=0.5$ and bare band dispersions as given in the text.
The dashed lines show the boundary of the reduced Brillouin zone of
the antiferromagnet with ${\vec Q}=(\pi/2,0)$.
a) Bare $c$ band.
b) Paramagnetic heavy Fermi liquid (FL), $V=0.3$.
c) Antiferromagnetic Fermi liquid (AF), with
c1) $V= 0.29$, $M_s = 10^{-4}$, $m_s = 10^{-5}$ (no changes occur for $M_s,m_s\to 0$),
c2) $V= 0.01$, $M_s = 0.1$, $m_s = 10^{-2}$.
From b) to c1), the system undergoes a standard SDW transition,
whereas c2) is close to a ``local-moment'' antiferromagnet
[i.e. obtained by backfolding of the bandstructure from a),
together with a fully gapped $f$ band].
Note that c) has two completely filled bands after backfolding.
}
\label{fig:fs}
\end{figure}

As example we focus on a
quasi-one-dimensional situation of a 2d array of coupled chains,
where the AF phase has a $4\times 1$ unit cell,
i.e. a collinear period-4 AF order in each chain.
In Fig.~\ref{fig:fs} we show the Fermi surface evolution and the band structures
for parameters
$\epsilon_{\vec k}    = - 2 \cos k_x + \cos (2 k_x) - (\cos k_y)/ 30$,
$\epsilon_{{\vec k}f} = 0.2 [\cos k_x - \cos (2 k_x)] - (\cos k_y)/50$,
$n_c = 0.5$, $n_f = 1$,
and different values of $V$, $M_s$, $m_s$.
The bare $c$ Fermi surface is shown in Fig.~\ref{fig:fs}a.
Backfolding into the AF zone results in two partially filled bands --
this is the FS structure of the local-moment antiferromagnet,
where the $f$ subsystem is gapped (Fig.~\ref{fig:fs}c2).
On the other hand, the paramagnetic large-FS state
for sizeable $V$ is shown in Fig.~\ref{fig:fs}b.
Entering the AF phase via an SDW transition
leaves us with two full and two partially filled bands,
Fig.~\ref{fig:fs}c1.
Importantly, this topology is identical to the
one in Fig.~\ref{fig:fs}c2.
Together with the continuity arguments in Sec.~\ref{sec:adiabatic},
this constitutes a proof that a continuous
evolution from itinerant (Fig.~\ref{fig:fs}c1) to
local-moment (Fig.~\ref{fig:fs}c2) antiferromagnetism is possible.

A few remarks are in order:
(i)
Due to the 1d geometry, there are no closed orbits,
and the large Fermi surface of FL does not intersect the AF zone
boundary (Fig.~\ref{fig:fs}b).
As a result, there is no Landau damping of the AF order parameter
at the SDW transition.
However, both features can be easily changed by including a second $c$ band
with a closed FS intersecting the AF zone boundary.
(Heavy-fermion metals often feature a complicated
band structure with numerous bands.)
If this band is only weakly hybridized with the $f$ band,
then its Fermi surface will not differ in topology between
the itinerant and local-moment limits.
We shall give an additional example in Appendix~\ref{app:ex2}
which displays both closed orbits and Landau damping.
(ii)
For an AF unit cell with $N$ a half-integer multiple of 4 ($N=2,6,\cdots$),
topological reasons require that the Fermi surface crosses at least one
van-Hove singularity between the itinerant and local-moment limits
(see Appendix~\ref{app:crit}).


\section{Renormalization group and fixed points}
\label{sec:rg}

In this section, we discuss the fate of Kondo screening in the framework
of the renormalization-group (RG) treatment of the Kondo coupling
between local moments and conduction electrons.
In this language, Kondo screening (in the paramagnetic phase)
is signaled by a runaway flow to strong coupling of
the Kondo interaction.

What happens in the antiferromagnetic phase?
Ref.~\onlinecite{yama} presented an RG calculation starting from
a local-moment antiferromagnet, with the result that the Kondo
coupling is an exactly marginal perturbation to the fixed point
of decoupled $c$ and $f$ electrons (dubbed \afs).
This qualitative difference to the paramagnetic case was used
to conclude that \afs\ and the Kondo-screened \afl\ situation
represent distinct {\em phases}, and a phase transition has to separate
the two.

For comparison, let us review the well-studied problem of a
single Kondo impurity in a magnetic field.\cite{hewson}
Here, no zero-temperature phase transition occurs as function
of the field, i.e., all observables evolve smoothly from
$B=0$ to $B\to\infty$.
For $B>0$, the low-energy theory takes the form of a spin-dependent
potential scatterer, with phase shifts varying continuously
as a function of the field.
In the RG language, this situation corresponds to a {\em line}
of fixed points, parameterized by the phase shifts.
This line connects the fixed points of the screened zero-field impurity
and the fully polarized impurity.
The topology of the RG flow implies the existence of an exactly
marginal operator -- for a finite fixed field,
this is the Kondo coupling $J$ itself.
Importantly, the distinct RG flow of $J$ for
zero field and finite field does {\em not} imply the
existence of a quantum phase transition.

Adapting this knowledge to the Kondo lattice,
we believe that the RG calculation of Ref.~\onlinecite{yama} is correct,
but does not imply the necessity for a phase transition
inside the antiferromagnetic phase
(if the latter is reached via a SDW transition).
We can offer two lines of arguments:
(a) Within DMFT, the antiferromagnetic phase of the Kondo lattice is
mapped onto a single Kondo impurity in a field,
hence the knowledge sketched above can be carried over.\cite{varma}
(b) Beyond DMFT, we know that both \afl\ and \afs\ are
Fermi-liquid phases, which generically display a set of marginal
operators (corresponding e.g. to band-structure parameters).
As the RG expansion in Ref.~\onlinecite{yama} is performed
around a Fermi liquid, finding marginal operators is no surprise.
(This is different from an expansion starting from decoupled moments
in the paramagnetic Kondo lattice -- this is a non-Fermi liquid due
to the degenerate local-moment states.)
In fact, it is straightforward to check that all Fermi-liquid renormalizations
are generated in the expansion of Ref.~\onlinecite{yama}.
Hence, no qualitative differences exist between the local-moment and
itinerant AF regimes.

In summary, the RG result of Ref.~\onlinecite{yama} is not
in contradiction with itinerant and local-moment AF regimes
being continuously connected.
Instead, the AF phase should be interpreted as family of RG
fixed points of Fermi-liquid type.
Depending on band structure details,
topological transitions may occur along a path from itinerant to local-moment AF,
but this is not required, see Sec.~\ref{sec:mf}.


\section{Slave particles, gauge fields, and phase diagram}
\label{sec:gauge}

A popular description of the low-temperature Kondo lattice physics
is based on a representation of the local moments
by auxiliary fermions and a decoupling of the Kondo interaction
by slave-boson fields $b_i$.
Fluctuation effects are captured via a compact U(1) gauge field.
An additional inter-moment exchange interaction $J_H$ is decoupled using a non-local
field $\chi$, such that the low-energy action takes the form:\cite{senthil2}
\begin{eqnarray}
\label{sbact}
{\cal S} &=& \int {\rm d}\tau
\Big\{
\sum_k \bar{c}_k (\partial_\tau -
\epsilon_k)c_k
+ \sum_i \bar{f}_i(\partial_\tau - ia_0)f_i \nonumber \\
&-& \sum_i \big(b_i\bar{c}_i f_i + {\rm c.c.} - \frac{4|b_i|^2}{J} \big) \nonumber\\
&-& \sum_{\langle ij\rangle} \big[
\chi_{ij} \left(e^{ia_{ij}}\bar{f}_i f_{j} + \mbox{c.c.} \right) - \frac{4|\chi_{ij}|^2}{J_H}\big]
\Big\}.
\end{eqnarray}
Here, the time component $a_0$ of the gauge field implements the
occupation constraint for the $f$ fermions, whereas the space component
$a_{ij}$ represents phase fluctuations of the decoupling parameters.

In the saddle-point approximation, gauge-field fluctuations are ignored.
This mean-field approach has well-known deficiencies, such
as an artificial breaking of the internal gauge symmetry
in the FL regime where $\langle b\rangle\neq 0$,
leading to an artificial finite-temperature phase transition.
These deficiencies are cured once the gauge-field physics is taken
into account.
After integrating out the fermions, the effective theory is given
by a compact U(1) gauge field coupled to the charged scalar $b$.
The Fermi-liquid phase corresponds to the Higgs phase of the
gauge theory; due to the compactness of the U(1) gauge field,
no finite-temperature transition occurs.
Most importantly, for this gauge theory, the confined and the Higgs phase
are smoothly connected, i.e. identical\cite{compact} (and hence FL-like).
The gauge theory also has a deconfined (or Coulomb) phase, characterized by
topological order.
Without magnetism, the result is the fractionalized Fermi liquid FL$^\ast$.
No Kondo screening occurs here, i.e., $b=0$ at the mean-field level.
The transition from FL to FL$^\ast$ is the Kondo-breakdown transition
advocated in Refs.~\onlinecite{senthil,senthil2}, which can also
be interpreted as a Mott transition of the $f$ electron subsystem.\cite{pepin}
Adding antiferromagnetic order to the FL$^\ast$ phase results
in a AF$^\ast$ phase -- this is a fractionalized antiferromagnet.\cite{senthil2}

\begin{figure}[!t]
\epsfxsize=2.7in
\epsffile{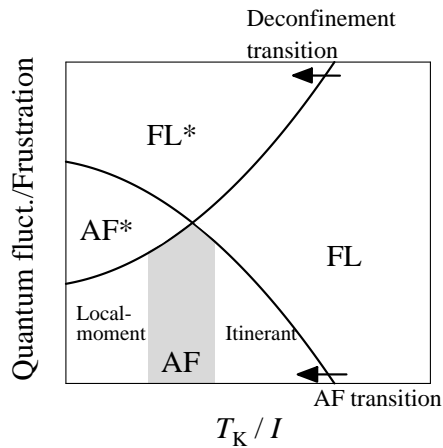}
\caption{
Schematic zero-temperature phase diagram of the Kondo-Heisenberg
model ${\cal H}_{\rm KLM}+{\cal H}_{\rm I}$,
as function of the dimensionless ratio $T_{\rm K}/I$
(formed from Kondo and inter-moment exchange energy scales)
and the amount of quantum fluctuations (e.g. due to frustration).
The solid lines show continuous phase transitions, associated with
the onset of antiferromagnetism and deconfinement/topological order.
In the absence of antiferromagnetism, the deconfinement transition
is equivalent to a Kondo breakdown or $f$ Mott transition.
The shaded area denotes the crossover from
itinerant to local-moment AF;
in this region FS-topology-changing transitions {\em may} occur,
depending on the type of magnetic order and the FS
topology, see Appendix~\ref{app:crit}.
(The structure of the phase boundaries follows from
Refs.~\onlinecite{senthil,senthil2}.)
A scenario of deconfined criticality modifies the phase
diagram,\cite{senthil3} see text.
}
\label{fig:pd}
\end{figure}

The zero-temperature phase diagram from this discussion is in Fig.~\ref{fig:pd}.
The smooth connection between Higgs and confined phases of the gauge theory
allows us to conclude that, in the presence of antiferromagnetism,
there is only a single {\em conventional} phase (AF).
Inside the AF phase, a transition towards the fractionalized AF$^\ast$ phase
is possible, accompanied by the onset of topological order.
Such a transition may be driven by increasing quantum fluctuations or
magnetic frustration.\cite{frust_sl}
However, it is usually assumed that local-moment antiferromagnetism
in heavy-fermion metals is conventional, such that the
AF$^\ast$ phase is unlikely to be realized.\cite{afstar}
(Mean-field theories display a zero-temperature transition
where $b$ vanishes upon increasing the magnetic order parameter.
This transition -- which may be interpreted as the AF--AF$^\ast$ transition --
does not coincide with a possible Lifshitz transition.)

Finally, we note that a Kondo-breakdown transition from FL to AF
in Fig.~\ref{fig:pd} is only possible with fine-tuning
via the multicritical point.
In contrast, in the scenario of deconfined criticality\cite{deconf} proposed for the
Kondo lattice,\cite{senthil3}
the FL$^\ast$ phase is assumed to display a secondary instability
toward antiferromagnetism (accompanied by confinement).
Then FL$^\ast$ (together with AF$^\ast$) disappears from the $T=0$ phase diagram,
rendering a Kondo-breakdown transition from FL to AF possible without fine-tuning.
The RG flow near this transition displays multicritical behavior.\cite{deconf}


\section{Conclusions}

We have argued that local-moment and itinerant antiferromagnetism
in heavy-fermion compounds are not necessarily distinct phases.
To this end, we have demonstrated adiabatic continuity along the path
(i)   itinerant Kondo-lattice AF,
(ii)  non-interacting two-band system with small exchange field and strong hybridization,
(iii) non-interacting two-band system with large exchange field and vanishing hybridization,
(iv) local-moment Kondo-lattice AF.
Here, the connection (i)--(ii) follows from the Fermi-liquid
properties of the heavy FL,
the connection (ii)--(iii) was established using an explicit
example for a continuous Fermi-surface evolution in Sec.~\ref{sec:mf},
and the connection (iii)-(iv) builds on the continuity between
antiferromagnetic band and Mott insulators in one-band models,
together with the Fermi-liquid properties of the local-moment
antiferromagnet.
This provides a proof of principle that a continuous evolution from
itinerant to local-moment antiferromagnetism is possible,
without intervening topological or other quantum phase transitions.

These arguments do not exclude the existence of (continuous or first-order) Lifshitz or van-Hove
transitions within the antiferromagnetic phase.
However, those transitions depend on band structure and topology and cannot
be associated with the breakdown of Kondo screening in the magnetic phase:
The concepts of quasiparticles and Fermi surfaces remain well-defined across
such transitions.\cite{assaad}
Our reasoning, which rests on the broken translational symmetry in the AF phase,
applies similarly to other variants of translational symmetry breaking in
the local-moment regime of Kondo lattices,
e.g. the formation of valence-bond solids.

Let us briefly consider the implications for the much discussed Kondo-breakdown
scenario for the antiferromagnetic quantum critical point.
Our arguments imply that the two distinct quantum phase transitions,
namely conventional SDW and Kondo breakdown,
connect the {\em same} phases,
namely a paramagnetic and an antiferromagnetic Fermi-liquid metal.
Then, it is clear that measurements only taken {\it inside} the antiferromagnetic
phase (even close to the critical point) do not allow to draw sharp
conclusions about the nature of the quantum phase transition.
Only (i) the finite-temperature quantum critical behavior
and (ii) the evolution of observables at low temperatures {\em across}
the quantum phase transition can distinguish between
the transition being of conventional SDW type or of Kondo-breakdown type.
In the latter case, e.g. a jump in the zero-temperature limit of
the Hall coefficient\cite{coleman} across the quantum phase transition
can be expected.
(Such a jump also occurs at a first-order transition between FL and AF,
as in the theory work of Ref.~\onlinecite{ogata},
however, there will be no finite-temperature quantum critical region,
but instead the first-order behavior will continue to finite $T$.)

On the experimental side, \CeNi\ appears to fit into the standard SDW
transition scenario, whereas \CeAu\ and \YbRhSi\ have been discussed as candidates
for AF transitions accompanied by the breakdown of the Kondo
effect.\cite{hvl}
In particular, Hall effect measurements on \YbRhSi\
support a jump in the Hall coefficient across the QPT.\cite{paschen}
Recent substitution experiments\cite{gegpriv} in \YbRhSi\ gave indications
for a separation of the magnetic and Kondo-breakdown transition signatures
(e.g. upon replacing Rh with Ir or Co),
opening the exciting opportunity to study the global phase diagram in
more detail.
In line with our arguments, we predict the signatures of the Kondo breakdown
to be weakened or smeared inside the antiferromagnetic phase even in the
low-temperature limit.
For \CeAu, we note that several transport experiments inside the AF phase\cite{hvl98,wilhelm}
could be nicely explained by a competition between AF order and Kondo screening --
this may be taken as evidence {\em against} a Kondo breakdown scenario for the
magnetic QPT in this material.
More detailed studies (e.g. Hall effect under pressure) are desirable.

An interesting case is that of CeRhIn$_5$.\cite{park,shishido}
Under pressure, de-Haas-van-Alphen measurements detected a change in the Fermi
surface properties at a critical pressure of $p_c \approx 2.3$ GPa.\cite{shishido}
In zero field, this system displays an intricate pressure-driven
interplay of antiferromagnetism and superconductivity.
However, the experiment was performed in fields up to 17 T,
where superconductivity is suppressed, but antiferromagnetism
is believed to persist for $p<p_c$.
Thus, the experimental data may be consistent with a Kondo-breakdown
transition upon lowering $p$, which occurs concomitantly
with the onset of AF order.
Evidence for a transition inside an AF phase,
with a Fermi surface reconstruction,
has been recently found\cite{goh} in CeRh$_{1-x}$Co$_{x}$In$_{5}$.
This transition, however, is strongly first order and also accompanied by
a change in the magnetic structure.

Last not least, we note that metallic spin-liquid behavior
has been observed in the geometrically frustrated Kondo lattice
compound Pr$_2$Ir$_2$O$_7$,\cite{met_sl}
rendering it a candidate for the FL$^\ast$ phase.
Here, experimental efforts to drive a transition towards FL
[by doping or (chemical) pressure] seem worthwhile.


\acknowledgments

I thank T. Senthil, Q. Si, S. Yamamoto, and especially A. Rosch for enlightening discussions,
and S. Sachdev and T. Senthil for collaborations on related work.
This research was supported by the DFG through
SFB 608 and the Research Unit FG 960 ``Quantum Phase Transitions''.


\appendix

\section{Band structure of itinerant antiferromagnets}
\label{app:band}

In this appendix, we highlight the distinctions between collinear and non-collinear AF
regarding band degeneracy.
Those are relevant for the possible existence of FS topological transitions
inside the AF phase of Kondo lattices, to be discussed in Appendix~\ref{app:crit}.

Consider the local-moment electrons only, with one electron per paramagnetic unit cell
and a dispersion $\epsilon_{\vec k}$,
in the presence of commensurate antiferromagnetic order described by a single
ordering wavevector $\vec Q$.
The electrons move in a mean field given by
${\vec h}(\vec r_i) = {\rm Re}[{\vec M} \exp(i \vec Q \cdot \vec r_i)]$,
where $\vec M$ is a complex vector encoding the order parameter, such
that e.g. spiral order has $\vec M = \vec M_1 + i\vec M_2$ with real $\vec M_1$ and $\vec M_2$
obeying $\vec M_1\cdot \vec M_2 = 0$.
For a unit cell with $N$ sites, $N\vec Q$ is a reciprocal lattice vector,
and the band structure is given by the eigenvalues of the $2N\times 2N$ matrix
\begin{equation}
\begin{pmatrix}
  \epsilon_{\vec k} & 0                 & M_{\up\up}               & M_{\up\dn}               & 0    & 0 & \ldots \\
  0                 & \epsilon_{\vec k} & M_{\dn\up}               & M_{\dn\dn}               & 0    & 0 &  \\
  M_{\up\up}^\ast   & M_{\up\dn}^\ast   & \epsilon_{\vec k+\vec Q} & 0                        & M_{\up\up} & M_{\up\dn}  \\
  M_{\dn\up}^\ast   & M_{\dn\dn}^\ast   & 0                        & \epsilon_{\vec k+\vec Q} & M_{\dn\up} & M_{\dn\dn}  \\
  0                 & 0                 & M_{\up\up}^\ast          & M_{\up\dn}^\ast          & \epsilon_{\vec k+2\vec Q} & 0 &  \\
  0                 & 0                 & M_{\dn\up}^\ast          & M_{\dn\dn}^\ast          & 0                         & \epsilon_{\vec k+2\vec Q} &  \\
  \ldots            &                   &                          &                          &  &  & \ldots
\end{pmatrix}
\label{mat}
\end{equation}
where odd (even) rows and columns correspond to spin up (down) electrons,
and $M_{\sigma\sigma'} = \vec M \cdot \vec \tau_{\sigma\sigma'}$ with $\vec \tau$ the vector
of Pauli matrices. For collinear AF, $M_{\sigma\sigma'}$ can be chosen diagonal $\propto \tau_z$.

We are interested in the degeneracy of the $2N$ bands.
The simplest situation is collinear AF with even $N$:
Here, the up and down sectors decouple and are degenerate as the eigenvalues of \eqref{mat}
depend only on $|\vec M|^2$.
For non-collinear order with even $N$ the spin sectors mix, but the double degeneracy is
preserved, because there exists a translation operation which reverses all spins
(i.e. the combined action of time reversal plus translation is a symmetry
of the state).

For odd $N$, the spin degeneracy is in general lifted.
The only exception is the case of a purely imaginary $M_{\sigma\sigma'}$:
Here, one site within the unit cell (more generally, an odd number of sites) has
exactly zero magnetization.
(This situation requires an exotic spin liquid, further symmetry breaking,
or coupling to other degrees of freedom like Kondo screening of these moments.)

We conclude that, for both the cases of
(i) even $N$ and
(ii) odd $N$ with spin degeneracy lifted,
the itinerant antiferromagnet with one electron per site,
has {\em integer} band filling
(after taking into account backfolding and spin).
Then, in the presence of sufficiently large exchange fields
(i.e. band splitting) the magnet will be insulating.
This AF band insulator is adiabatically connected to an
AF Mott insulator.


\section{Criteria for the non-existence of FS topological transitions}
\label{app:crit}

As transitions with changes in the FS topology may always occur within Fermi-liquid phases,
we discuss under which circumstances topological transitions in the AF phase of
Kondo lattices are {\em not} required to occur:
An example without transitions was given in Sec.~\ref{sec:mf}.
The discussion will concentrate on the momentum-space volume and topology of
the occupied fermionic states of the $c$ plus $f$ electron system,
where we allow for $K$ different $c$ electron bands.
The strategy is to find conditions for the backfolded band structures
of the FL phase and the $c$ band(s) alone to be identical in topology --
then a smooth evolution from itinerant to local-moment AF is possible.

We denote the total occupied volumes of the partially filled bands
in the FL phase and the $c$ band(s) alone
with $K_d n_L$ and $K_d n_S$, respectively.
Hence,
$n_L \equiv (n_c+1)\,{\rm mod}\,2$ and
$n_S \equiv n_c\, {\rm mod}\,2$, in other words,
$n_L$ and $n_S$ differ by an odd integer.
After backfolding, both total occupied volumes (again of the partially filled bands)
need to be identical to avoid topological transitions in the AF phase.
The volumes can change by $2/N$ if bands after backfolding (i.e.
in the reduced Brillouin zone) are fully occupied (as in Fig.~\ref{fig:fs}b).
Hence, the number $F$ of completely filled {\em reduced} BZ must differ
by an odd multiple of $N/2$ between the FL phase and the $c$ bands alone,
$(F_L-F_S) 2/N = 1 \,{\rm mod}\,2$.

This condition can obviously not be met for odd $N$;
here ``large'' and ``small'' Fermi volume are only equivalent in the AF phase
after taking into account the broken spin degeneracy of the bands.
Thus, for odd $N$, at least one topological transition occurs inside the AF phase.

For even $N$, we now focus on the number $P$ of partially filled bands in the AF phase
(after backfolding)
where those $(d-1)$ dimensional surface areas of the reduced BZ boundary, which were
not part of the original BZ boundary, are fully occupied.
For the discussion we furthermore assume inversion symmetry.
$P$ receives {\em even} contributions from all bands of the paramagnetic phase,
except for situations where a reduced BZ is fully occupied:
Each of these cases give an {\em odd} contribution to $P$,
hence $P \equiv F\,{\rm mod}\,2$.
To avoid topological transitions, $P$ must be equal for the itinerant
AF (derived from FL) and the localized AF (derived from the $c$ bands alone),
$P_L=P_S$.
Hence, $F_L \equiv F_S\,{\rm mod}\,2$.
The above condition $F_L-F_S = (N/2)\,{\rm mod}\,N$ translates into the necessary condition
of $N$ to be a multiple of 4 for not having a topological transition
(in the presence of inversion symmetry).
An example with $N=4$, $K=1$, $P=2$ is in Fig.~\ref{fig:fs}.

Further considerations now have to include the $(d-2)$ dimensional ``edges'' of the BZ.
Here, we have found that at least in low-symmetry situations topological transitions
can be avoided.
One truly 2d example, which also includes closed electron orbits,
with $N=4$, $K=1$, $P=0$ is in Fig.~\ref{fig:fs2}.


\section{Fermi-surface evolution: Another example}
\label{app:ex2}

We provide an additional example for a continuous Fermi-surface
evolution within a model of effectively non-interacting electrons,
similar to Sec.~\ref{sec:mf}.

\begin{figure}[!t]
\epsfxsize=3.5in
\epsffile{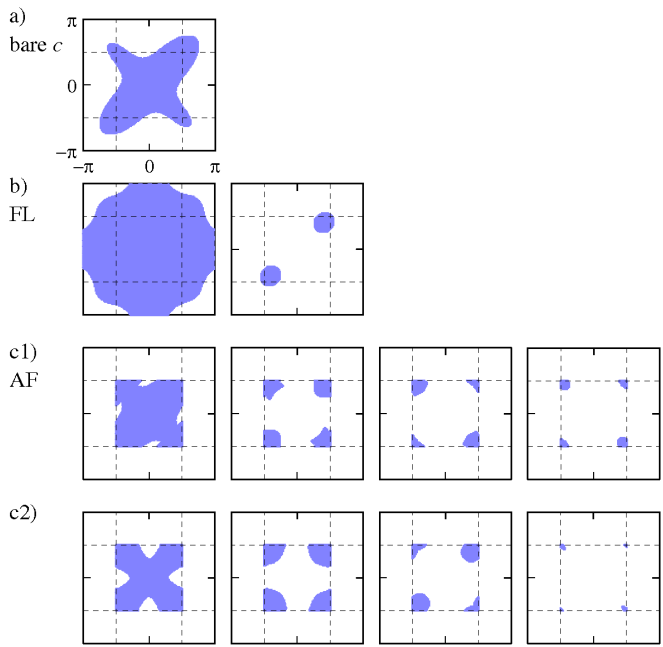}
\caption{
(Color online)
Fermi surfaces as in Fig.~\ref{fig:fs},
but now for a situation with a $2\times2$ AF unit cell
as described in the text.
a) Bare $c$ band.
b) Paramagnetic heavy Fermi liquid (FL), $V=0.45$.
c) Antiferromagnetic Fermi liquid (AF), with
c1) $V= 0.44$, $M_s = 10^{-4}$, $m_s = 10^{-5}$,
c2) $V= 0.01$, $M_s = 0.1$, $m_s = 10^{-2}$.
}
\label{fig:fs2}
\end{figure}

The model is defined on a 2d square with inequivalent diagonals
(equivalent to a rhomboid lattice).
The AF phase has $N=4$ with a $2\times 2$ unit cell,
originating from 4-sublattice co-planar AF order
characterized by two ordering wavevectors $\vec Q_1=(\pi,0)$ and
$\vec Q_2=(0,\pi)$, where $\vec Q_1$ ($\vec Q_2$) correspond
to a spin-density wave with spin polarization in $z$ ($x$) direction
with equal amplitude.
The model parameters are
$\epsilon_{\vec k}    = (\cos k_x - \cos k_y)^2 + (\cos 2k_x + \cos 2 k_y)/2 - \cos(k_x-k_y) + [(1-\cos k_x)^8+(1-\cos k_y)^8]/50$,
$\epsilon_{{\vec k}f} = -(\cos k_x + \cos k_y)/10 + [\cos(k_x-k_y) + \cos(k_x+k_y)]/20 + (\cos 2k_x + \cos 2 k_y)/20$,
$n_c = 0.65$, $n_f = 1$.

The Fermi surface evolution is shown in Fig.~\ref{fig:fs2}.
Backfolding of the FL bands (Fig.~\ref{fig:fs2}b)
results in two completely filled plus four partially filled bands
in the itinerant antiferromagnet (Fig.~\ref{fig:fs2}c1).
Again, these Fermi surfaces are topologically identical to those of the
local-moment antiferromagnet (Fig.~\ref{fig:fs2}c2), obtained from
backfolding the bare $c$ band structure of Fig.~\ref{fig:fs2}a.
Note that this example displays both closed electron orbits
and Landau damping of the order parameter.


\section{Incommensurate magnetism}
\label{app:incomm}

So far, the discussion applied to antiferromagnetism with a spatial period
being commensurate with the crystal lattice.
Incommensurate order is qualitatively different, at least at $T=0$
in an ideal crystal.
Here, the volume of the reduced Brillouin zone (after backfolding)
is zero, and the resulting band structure has quasi-crystalline properties,
with a hierarchy on infinitely many gaps.\cite{paschen,rosch}
Hence, the Fermi volume is no longer a well-defined concept.
Another way to think about incommensurate ordering is
to approach the incommensurate $\vec Q$ via a sequence of commensurate
$\vec Q$, with increasing size of the unit cell and decreasing size
of the Brillouin zone.
A small Brillouin zone implies that changes of the band structure
cause frequent topological transitions.
Then, the passage from itinerant to local-moment antiferromagnetism
in the incommensurate case will be accompanied by a dense sequence of
topological transitions -- which may be interpreted as adiabatic continuity.

In the presence of finite temperature or disorder, the small band gaps
will be smeared, and the structure effectively behaves as commensurate
(for fixed $T$). Then, our above consideration in the body of the
paper apply.



\begin{thebibliography}{}
\bibitem{rkkyfoot}
While the Kondo lattice model, Eq.~\eqref{eq:KLM}, contains
the physics of the indirect RKKY interaction between local
moments, it is convenient for the theoretical discussion
to introduce an (additional) explicit inter-moment interaction.
\bibitem{scfoot}
In principle, the FL phase can be unstable to
superconductivity at very low temperatures.
We shall ignore this here, i.e., we work at
low temperatures above the superconducting $T_c$.
\bibitem{affoot}
The main effects of collinear itinerant antiferromagnetism are the breaking
of SU(2) symmetry, band backfolding, and the opening of band gaps
at the boundary of the antiferromagnetic Brillouin zone.
These are correctly described in the standard saddle-point treatment,
where order is implemented via a staggered mean field.
Fluctuation effects, like collective modes of the ordered state,
are not captured.
However, those are unimportant for our discussion of adiabatic
continuity.
\bibitem{afstar}
Experimentally, AF$^\ast$ may be detected via broad continua in neutron
scattering or via its low-energy gauge-field excitations.\cite{senthil2}
\end{thebibliography}
\end{document}